\input harvmac
\input psfig
\newcount\figno
\figno=0
\def\fig#1#2#3{
\par\begingroup\parindent=0pt\leftskip=1cm\rightskip=1cm\parindent=0pt
\global\advance\figno by 1
\midinsert
\epsfxsize=#3
\centerline{\epsfbox{#2}}
\vskip 12pt
{\bf Fig. \the\figno:} #1\par
\endinsert\endgroup\par
}
\def\figlabel#1{\xdef#1{\the\figno}}
\def\encadremath#1{\vbox{\hrule\hbox{\vrule\kern8pt\vbox{\kern8pt
\hbox{$\displaystyle #1$}\kern8pt}
\kern8pt\vrule}\hrule}}
\def\underarrow#1{\vbox{\ialign{##\crcr$\hfil\displaystyle
 {#1}\hfil$\crcr\noalign{\kern1pt\nointerlineskip}$\longrightarrow$\crcr}}}
%
\overfullrule=0pt

%
\def\tilde{\widetilde}

\def\Z{{\bf Z}}

\def\S{{\bf S}}
\def\R{{\bf R}}
\def\CP{{\bf CP}}

\font\zfont = cmss10 

\def\bigone{\hbox{1\kern -.23em {\rm l}}}
\def\ZZ{\hbox{\zfont Z\kern-.4emZ}}

\Title{hep-th/0108165} {\vbox{\centerline{Anomaly Cancellation On
$G_2$-Manifolds}
\bigskip
\centerline{ }}}
\smallskip
\centerline{Edward Witten}
\smallskip
\centerline{\it School of Natural Sciences, Institute for Advanced Study}
\centerline{\it Olden Lane, Princeton, NJ 08540, USA}\bigskip

\medskip

\noindent
Smooth manifolds of $G_2$ holonomy, used to compactify $M$-theory
to four dimensions, give only abelian gauge groups without
charged matter multiplets.  But singular $G_2$-manifolds can give
abelian or nonabelian gauge groups with chiral fermions.  We
describe the mechanism of anomaly cancellation in these models,
using anomaly inflow from the bulk.  We also compare the anomaly
predictions to what has been learned by more explicit arguments
in some special cases.

\Date{August, 2001}
\newsec{Introduction}

\def\N{{\cal N}}
Compactification on a manifold $X$ of $G_2$ holonomy is a natural
framework for reducing eleven-dimensional $M$-theory to a
four-dimensional model with $\N=1$ supersymmetry. This type of
model is much harder to study than Calabi-Yau compactification,
because  Yau's theorem, which gives a useful criterion for
existence of Calabi-Yau metrics, has no analog for metrics of
$G_2$ holonomy.  Nonetheless, such metrics exist in many cases;
for an account of the existence proofs, see \ref\joyce{D. Joyce,
``Compact Manifolds Of Special Holonomy'' (Oxford University
Press, 2000).}.

A different kind of problem is that compactification on a large
and smooth manifold of $G_2$ holonomy -- the only case in which
supergravity is adequate -- gives a four-dimensional model with
abelian gauge fields only and no massless charged matter
multiplets. (The dimensional reduction on a manifold of $G_2$
holonomy has been worked out most fully in \ref\townsend{P.
Townsend and G. Papadopoulos, ``Compactification Of $D=11$
Supergravity On Spaces Of Exceptional Holonomy,'' hep-th/9506150.
Phys. Lett. {\bf B357} (1995) 472.}.) In fact, since a manifold
of $G_2$ holonomy has no continuous symmetries, gauge fields come
only from the dimensional reduction of the three-form field $C$.
Such gauge fields are abelian and, in the supergravity
approximation, which is valid for massless fields if $X$ is
smooth and large, do not couple to any charged fields at all.

To derive an interesting model of particle physics from a
manifold of $G_2$ holonomy, we therefore must allow singularities.
Some special cases of singularities of $G_2$ manifolds were
studied recently from different points of view \nref\acharya{B.
Acharya, ``On Realising ${\cal N}=1$ Super Yang-Mills In
$M$-Theory,'' hep-th/0011089.}%
\nref\amv{M. F. Atiyah, J. Maldacena, and C. Vafa, ``An
$M$-Theory Flop As A Large
$N$ Duality,'' hep-th/0011256.}%
 \nref\aw{M. F. Atiyah and E. Witten,
``$M$-Theory Dynamics On A
Manifold Of $G_2$ Holonomy,'' hep-th/0107177.}%
\nref\csq{M. Cvetic, G. Shiu, and A. M. Uranga, ``Three Family
Supersymmetric Standard-Like Models From Intersecting Brane
Worlds,'' hep-th/0107143, ``Chiral Four-Dimensional ${\cal N}=1$
Supersymmetric Type IIA Orientifolds
From Intersecting $D6$ Branes,'' hep-th/0107166.}%
\nref\newach{B. Acharya and E. Witten, to appear.}%
 \refs{\acharya -
\newach}, and some of these give nonabelian gauge symmetry and/or
chiral fermions.

In this paper, we will explore one route to obtain nonabelian
gauge symmetry.  In this approach, the generic singularities of
$X$ are codimension four $A-D-E$ orbifold singularities, which
give gauge symmetry.  Chiral fermions arise when the  locus of
$A-D-E$ singularities
 passes through isolated points at which $X$ has an isolated
conical singularity that is not just an orbifold singularity.
This approach to obtaining nonabelian gauge groups and chiral
fermions from a singular $G_2$-manifold can be motivated by
duality with Type IIA orientifolds such as those studied in \csq:
in those models, the gauge symmetry is carried by stacks of
branes, which lift to $A-D-E$ singularities in $M$-theory, and
the chiral fermions are supported at special points on the branes.
The same picture can also be motivated by duality with the
heterotic string, as will be explained elsewhere
\newach.  The heterotic string approach (as will be explained in \newach)
involves a $G_2$ analog of a familiar mechanism to obtain charged
hypermultiplets for Type IIA \ref\kv{S. Katz and C. Vafa,
``Matter From Geometry,'' Nucl. Phys. {\b B497} (1997) 146,
hep-th/9606086.}.

In section 2, we consider the case that $X$ is smooth except for
isolated conical singularities. The gauge group is then abelian,
coming from the $C$-field.  Gauge anomalies  can be used to
predict that chiral fermions (charged under the abelian gauge
group) must be present at conical singularities under a certain
topological condition.   The anomalies (which cancel by a
mechanism involving anomaly inflow from the bulk, rather as in
\ref\callanharvey{C. G. Callan, Jr. and J. A. Harvey, ``Anomalies
And Fermion Zero Modes On Strings And Domain Walls,'' Nucl. Phys.
{\bf B250} (1985) 427.})  give a constraint on the chiral
fermions.   In particular, for examples considered in \aw\ where
the topology has been worked out and spectra of chiral fermions
have been proposed, we show that these spectra agree with the
anomaly constraints.  The fact that isolated singularities of a
$G_2$-manifold can support charged chiral fermions is somewhat
analogous to the fact that charged hypermultiplets can be
supported at isolated singularities of a Calabi-Yau threefold in
Type IIA \ref\strominger{A. Strominger, ``Massless Black Holes
And Conifolds In String Theory,'' Nucl. Phys. {\bf B451} (1995)
96, hep-th/9504090.}.

In section 3, we incorporate the $A-D-E$ singularities.  Anomaly
considerations imply that chiral fermions must arise, under
certain conditions, if a singularity of type $A$ passes through
an isolated point at which $X$ has a (non-orbifold) conical
singularity.  Again, the results that come from anomalies can be
compared, in special cases, to results obtained in \aw. Anomalies
again cancel by a sort of inflow from the bulk.

Some rudimentary model-building observations based on this
mechanism for obtaining chiral fermions will appear elsewhere.

Apart from what is discussed in the present paper, another problem
of recent interest for which the cone on $\CP^3$ (discussed in
\aw\ and here) is relevant is the problem of a flop occurring in
the strongly coupled limit of the heterotic string on a Calabi-Yau
threefold $W$.  When such a flop occurs, the familiar
compactification manifold  $W\times \S^1/\Z_2$ of the strongly
coupled heterotic string is replaced by a more complicated
spacetime, studied in \ref\greenetal{B. R. Greene, K. Schalm, and
G. Shiu, ``Dynamical Topology Change In $M$ Theory,''
hep-th/0010207.}, with a singularity in the bulk that is a cone on
$\CP^3$. This might suggest generalizing our discussion to
include a $G$-flux, as in \greenetal, or generalizing the
discussion in \greenetal.

In this paper, we consider continuous gauge symmetries and local
anomalies only.  We do not attempt to analyze global anomalies or
anomalies in discrete gauge symmetries.

\newsec{Anomalies And Chiral Fermions From Isolated Conical Singularities}

First we consider the case that $X$ is smooth except for isolated
conical singularities.  Near such a singularity, the metric on
$X$ looks like \eqn\injo{ds^2=dr^2+r^2d\Omega^2,} where
$d\Omega^2$ is a metric on a six-manifold $Y$. The radial
variable $r$ is nonnegative, and the singularity is at $r=0$.

In general, we want to assume that there are isolated points
$P_\alpha\in X,\,\alpha=1,\dots, s$ at which $X$ has conical
singularities. Near $P_\alpha$, we assume that $X$ looks like a
cone on some six-manifold $Y_\alpha$.  If we excise from $X$
small open neighborhoods of the $P_\alpha$, we make a smooth
manifold-with-boundary $X'$; its boundary is the union of the
$Y_\alpha$.

First let us discuss the gauge group in $M$-theory on $X$. If $X$
is smooth, this can be determined by conventional Kaluza-Klein
reduction.  The gauge group (apart from discrete gauge symmetries
coming from symmetries of $X$)  is $H^2(X;U(1))$, coming from the
unbroken symmetries of the $C$-field.  The continuous gauge
symmetries in this situation can be described particularly
simply; if we take a basis $w_1,\dots,w_r$ of harmonic forms on
$X$ (where $r=b_2(X)$, the second Betti number of $X$), then
corresponding massless $U(1)$ gauge fields $A^{(i)}$ arise in
four dimensions by the ansatz \eqn\ixo{C=\sum_{i=1}^r
A^{(i)}\wedge w_i+\dots.}  To get the right global structure of
the gauge group $U(1)^r$, we normalize the $w_i$ to be generators
of $H^2(X;\Z)$.

In case $X$ has isolated conical singularities, we propose that
the gauge group $L$ is $H^2(X';U(1))$.  This is the answer one
gets if the harmonic forms $w_i$ are required, near a singular
point that is a cone on $Y_\alpha$, to be ``pullbacks'' from
$Y_\alpha$, that is, arbitrary  harmonic forms on $Y_\alpha$ but
independent of the radial variable $r$. The proposal that
$L=H^2(X';U(1))$ agrees with examples considered in \aw\ and will
lead, as we will see, to an elegant general picture for anomaly
cancellation.\foot{It might be that something new would happen if
the first Betti number of $Y_\alpha$ is nonzero (no examples are
known of  conical $G_2$ singularities with this property).
Perhaps then we should allow harmonic forms that near a
singularity look like $dr$ times a one-form on $Y_\alpha$.}

We will use anomalies to obtain some information about charged
chiral fermions that must be supported at the points $P_\alpha$.
We consider first purely gauge anomalies (as opposed to mixed
gauge-gravitational anomalies).  Denoting the curvature of $C$ as
$G=dC$, the anomalies come from the $C G^2$ interaction of
eleven-dimensional supergravity.  When properly normalized (so
that the periods of the $G$-field are multiples of $2\pi$), this
interaction, which we will call  $I$, is such that\foot{Together
with gravitational corrections, $I$ is defined mod $2\pi$ and
hence $I/2\pi$ is defined mod 1, as explained in \ref\witten{E.
Witten, ``On Flux Quantization In $M$-Theory And The Effective
Action,'' hep-th/9609122, J. Geom. Phys. {\bf 22} (1997) 1.}.
(This assertion ignores a subtlety about the Rarita-Schwinger
determinant  that will not be important in the present paper.)}
 \eqn\tubu{{I\over 2\pi}={1\over
6\cdot (2\pi)^3}\int_M C\wedge G\wedge G.} Here $M$ is spacetime,
which for our purposes is $\R^4\times X$. However, this
supergravity formula is really only valid away from the
singularities of $X$. So we will really carry the integral only
over $M'=\R^4\times X'$.   This will give an anomalous result,
and we will cancel the anomaly with a suitable assumption about
the nature of the physics at the singularities.

To see the anomaly, we consider a gauge transformation of $C$, by
$C\to C+ d\epsilon$, with $\epsilon$ a two-form.  We want to prove
gauge-invariance by integrating by parts and using the fact that
$dG=0$.  In doing this, an anomaly will arise at the
singularities.  In fact, under $C\to C+d\epsilon$, since the
boundary of $M'$ is $\bigcup_\alpha \R^4\times Y_\alpha$, $I$
changes by \eqn\gurf{{\delta I\over 2\pi} = -{1\over 6\cdot
(2\pi)^3}\sum_\alpha \int_{\R^4\times Y_\alpha}\epsilon \wedge G
\wedge G.} Now we consider the case that near a singular point,
the ansatz \ixo\ is valid; we write $F^{(i)}=dA^{(i)}$ for the
field strengths of the four-dimensional fields; and we take
\eqn\wetak{\epsilon =\sum_i\epsilon^{(i)}w_i,} where
$\epsilon^{(i)}$ are functions on $\R^4$.   Under these
assumptions, the contribution of the $\alpha^{th}$ singular point
to the anomaly is \eqn\bego{{\delta_\alpha I\over 2\pi} = -{1\over
6\cdot (2\pi)^3} \int_{\R^4} \sum_{i,j,k} \epsilon^{(i)}\wedge
F^{(j)}\wedge F^{(k)}\int_{Y_\alpha} w_i\wedge w_j\wedge w_k.}

Anomaly cancellation must hold locally on $M$; hence there must
be some additional phenomenon supported at $\R^4\times P_\alpha$
that cancels the anomaly.  We will suppose that this additional
phenomenon takes the form of charged chiral superfields
$\Phi^\sigma$,  of charges $q^{\sigma}_i$. Here $\sigma $ takes
values in a set $T_\alpha$ that depends on $\alpha$, so the $q$'s
depend on $\alpha$ though this is not made explicit in the
notation.  Since anomalies in four-dimensional gauge theory are
derived from the six-form piece of $\exp(F/2\pi)$, which is
$(1/6) (F/2\pi)^3$, to cancel the anomaly \bego\ we require that
\eqn\rego{\sum_{\sigma\in T_\alpha}
q^{\sigma}_iq^{\sigma}_jq^\sigma_k=\int_{Y_\alpha}w_i\wedge
w_j\wedge w_k.}

Thus in particular, for $Y_\alpha$ such that the right hand side
is not identically zero, there must be massless chiral fermions
supported at $P_\alpha$.  Now let us verify anomaly cancellation.
Anomaly cancellation means that the anomalies of the massless
chiral fermions add up to zero, after summing over $\alpha$. That
is, we want \eqn\hego{\sum_\alpha\sum_{\sigma\in
T_\alpha}q^\sigma_iq^\sigma_j q^\sigma_k=0.} In virtue of \rego,
this follows from the fact that
$\sum_\alpha\int_{Y_\alpha}w_i\wedge w_j\wedge
w_k=\int_{X'}d(w_i\wedge w_j\wedge w_k)=0$, as $dw_i=0$ for all
$i$.

Now let us verify that our result for the local anomaly agrees
with the spectra found in \aw\ in special cases.  For $Y=\CP^3$,
the second Betti number is 1, and the second cohomology group is
generated by a single two-form $w$ with $\int_Y w^3=1$. Hence, we
expect a chiral spectrum with charges $\sigma$ such that
$\sum_\sigma (q^\sigma)^3=1$.  This agrees with the claim in \aw\
that there is a single chiral fermion of charge 1. For
$Y=SU(3)/U(1)^2$, the gauge group is $U(1)^2$, conveniently
embeddable in $U(1)^3 $ in such a way that the charges are
$(1,-1,0)$, $(0,1,-1)$, and $(-1,0,1)$.  In this basis, the
nonzero elements of the anomaly polynomial $d_{ijk}=\sum_\sigma
q^\sigma_iq^\sigma_jq^\sigma_k$ are (up to permutations of the
indices) $d_{i,i,i-1}=-d_{i,i,i+1}=1$.  This agrees with the
intersection form of $SU(3)/U(1)^2$.  (A key example in
\refs{\acharya - \aw} was the case $Y=\S^3\times \S^3$, but that
case is not very interesting for the present paper as the second
Betti number of $Y$ vanishes.)

Turning things around, we have canceled the anomalies from chiral
fermions that ``live'' at the singularity using a sort of anomaly
inflow from the $C$-field (roughly along the lines of
\callanharvey)
 and without invoking a Green-Schwarz mechanism. The fact that anomaly
 inflow is the key mechanism is not surprising, since this is the case for
 intersecting branes in Type II superstrings
\ref\plok{M. B. Green, J. A. Harvey, and G. Moore, ``$I$-Brane
Inflow And Anomalous Couplings On $D$-Branes,'' Class. Quant.
Grav. {\bf 14} (1997) 47, hep-th/9605033.}, and such brane
intersections are in some cases dual to isolated singularities of
a $G_2$ manifold. The class of models considered here has no
Green-Schwarz mechanism in bulk. It might be that  the local
degrees of freedom at some singularities are more complicated than
we have assumed (there might be a nontrivial conformal field
theory at a singularity, for example), and perhaps there are some
cases in which a description using a local Green-Schwarz
mechanism at a singularity is useful.

\bigskip\noindent{\it Mixed Gauge-Gravitational Anomalies}

Now let us consider the analogous mechanism for mixed
gauge-gravitational anomalies.  (There are no purely
gravitational anomalies in four dimensions.) The coupling
analogous to \tubu\ is the gravitational contribution to the
Chern-Simons coupling, \eqn\nubu{{I'\over
2\pi}=-{1\over48}\int_M{C\over 2\pi}\wedge
\left(p_2-p_1^2/4\right).} Here by $p_1$ and $p_2$ we mean the
differential forms (polynomials in the Riemann tensor) that
represent the Pontryagin classes $p_i$.

Now, under $C\to C+d\epsilon$, we get an additional anomaly:
\eqn\piko{ {\delta I'\over 2\pi}={1\over 48 }
\sum_\alpha\int_{\R^4\times Y_\alpha} {\epsilon\over 2\pi} \wedge
(p_2-p_1^2/4).} To extract the four-dimensional gravitational
anomaly, we want to evaluate this for fluctuations in the metric
of $\R^4$ that preserve the product form $\R^4\times X$ (but of
course not the flatness of $\R^4$).  For this purpose, if $p_1'$
and $p_1''$ are the four-forms representing the first Pontryagin
classes of $\R^4$ and $X$, respectively, we can take $
p_1=p_1'+p_1''$ and $p_2=p_1'\wedge p_1''$.  Also expanding
$\epsilon$ as in \wetak,  the local contribution to the anomaly
from the $\alpha^{th}$ singularity is  \eqn\riko{{\delta_\alpha
I'\over 2\pi}={1\over 96 }  \int_{\R^4}{\epsilon^{(i)}\over 2\pi}
p_1' \int_{Y_\alpha}w_i\wedge p_1''.}

Let us work out the anomaly that is expected due to chiral
fermions at $Y_\alpha$ of charges $q^\sigma_i$, $\sigma\in
T_\alpha$.  The anomaly for a chiral fermion of charge 1 in four
dimensions is derived from the six-form \eqn\turnigo{ {1\over
6}\left({F\over 2\pi}\right)^3-{F\over 2\pi}{p_1'\over 24}.} So
to cancel the local anomaly found in \riko, the chiral multiplets
supported at $Y_\alpha$ must have charges such that
\eqn\xiko{\sum_{\sigma\in T_\alpha} q^\sigma_i = {1\over
4}\int_{Y_\alpha} w_i\wedge p_1''.} Anomaly cancellation is now
established just as for the purely gauge anomalies:
\eqn\ancan{\sum_\alpha\sum_{\sigma\in T_\alpha}q^\sigma_i =
{1\over 4}\sum_\alpha \int_{Y_\alpha}w_i\wedge p_1''={1\over 4}
\int_{X'} d(w_i\wedge p_1'') = 0.}

Again, we can compare to the cases considered in \aw.  Suppose
that one of the $Y_\alpha$'s is a cone on $\CP^3$. The relevant
gauge group is $U(1)$, associated as above with a two-form $w$
with $\int_{\CP^3}w^3=1$. For $\CP^3$, we have $c_1=4w$,
$c_2=6w^2$, and $p_1 = c_1^2-2c_2=4w^2$. So
$(1/4)\int_{\CP^3}w\wedge p_1''=1$, and \xiko\ is compatible with
the expectation that there is precisely one chiral multiplet with
$q=1$.

For the other example, $Y=SU(3)/U(1)^2$, things are more
trivial.  There is a triality symmetry, exploited in \aw, which
ensures that the charge generators are traceless (this is clear
from the expressions for the charge vectors given above), and
likewise ensures that $p_1''=0$.  (The latter statement holds
because $H^4(Y;\Z)$ is a rank two lattice that is ``rotated'' by
the triality symmetry, in such a way that there are no nonzero
invariant vectors.)

\newsec{Incorporating Nonabelian Gauge Symmetry}

\nref\landlop{K. Landsteiner and E. Lopez, ``New Curves From Branes,''
Nucl. Phys. {\bf B516} (1998) 273, hep-th/9708118.}%
\nref\witen{E. Witten,
``Toroidal Compactification Without Vector Structure,'' JHEP {\bf 9802:006} (1998)
hep-th/9712028. }%
\nref\deboer{J. de Boer, R. Dijkgraaf, K. Hori, A. Keurentjes, J.
Morgan, D. R. Morrison, and S. Sethi,  ``Triples, Fluxes, And
Strings,'' hep-th/0103170.}%
 $M$-theory at a simple $A-D-E$ singularity generates gauge symmetry
of type $A-D-E$. (There also are singularities of type $D$ and
$E$ with gauge symmetry of reduced rank \refs{\landlop - \deboer},
but we will not consider them here.) If $X$ is a manifold of
$G_2$ holonomy with an $A-D-E$ singularity, then, as $X$ has
dimension 7 and the $A-D-E$ singularity has codimension four, the
singularity is supported  on a three-manifold $Q\subset X$.  We
use the term ``manifold'' somewhat loosely; like $X$ itself, $Q$
may have singularities.  $Q$ is somewhat analogous to a
supersymmetric three-cycle in $X$; in fact, locally, near $Q$ and
away from non-orbifold singularities, $X$ is a quotient $X=\tilde
X/\Gamma$ where $\Gamma$ is a finite group, and $Q$ is a
supersymmetric three-cycle in $\tilde X$ that is the fixed point
set of $\Gamma$.

The low energy gauge theory -- away from singularities -- is
supersymmetric $A-D-E$ gauge theory on $\R^4\times Q$.  We want
to understand the contributions of singularities, and in
particular we want to know what singularities support chiral
multiplets in complex representations of the gauge group. We
suppose that the singularities  are either isolated singularities
of $Q$, or points at which $Q$ is smooth but has a normal bundle
with a singularity worse than the generic $A-D-E$ singularity.
(In fact, the known examples and additional ones that will be
discussed in \newach\ are of the second type.) In this section, we
will use anomalies to give a constraint on chiral fermions from
singularities.

To define the $A-D-E$ singularity, we start with $\R^4$, acted on
by $SO(4)\cong SU(2)_L\times SU(2)_R$.  Then we pick a discrete
subgroup $\Gamma$ of $SU(2)_R$, and define the $A-D-E$ singularity
as the quotient $\R^4/\Gamma$.  The $A-D-E$ singularity  has as a
symmetry group $SU(2)_L\times \Lambda$, where $\Lambda$ is the
subgroup of $SU(2)_R$ that conjugates $\Gamma$ to itself (thus,
$g \Gamma g^{-1}= \Gamma$ for $g\in \Lambda$).

\nref\mn{J. Maldacena and C. Nunez, ``Towards The Large $N$ Limit Of Pure
${\cal N}=1$ Super Yang-Mills,'' Phys. Rev. Lett. {\bf 86} (2001) 588, hep-th/0008001.}%
\nref\manya{B. Acharya, J. P. Gauntlett, and N. Kim, ``Fivebranes Wrapped On
Associative Three-Cycles,'' hep-th/0011190.}%
In general, a family of $A-D-E$ singularities, over a base $B$,
can be ``twisted'' by an arbitrary $SU(2)_L\times \Lambda$
bundle. The case of main interest to us is that $B=\R^4\times Q$
(where $\R^4$ is four-dimensional Minkowski space).  The twisting
by $SU(2)_L$ can be described very directly: the condition that
$X$ has $G_2$ holonomy identifies the $SU(2)_L$ connection with
the Riemannian connection of $Q$. However, $G_2$ holonomy does
not determine how the normal bundle is twisted by $\Lambda$, and
we must examine this.\foot{To get a rough idea of the group
theory here (see \refs{\mn,\manya} for a more detailed discussion
of analogous problems that depend upon the same group theory),
$G_2$ contains the group $SU(2)_L\times SU(2)_R$, with the ${\bf
7}$ of $G_2$ transforming as $({\bf 3},{\bf 1})\oplus ({\bf
2},{\bf 2})$.  The $({\bf 3},{\bf 1})$ is the tangent space to
$Q$ and the $({\bf 2},{\bf 2})$ is the normal space to $Q$ (before
dividing by $\Gamma$ to make an orbifold). The last statement
shows that the group $SU(2)_L$ that acts on the normal bundle is
the same as the group that acts on the tangent space to $Q$ --
which is why the $SU(2)_L$ connection on the normal bundle is the
spin connection.  It also shows that an arbitrary twisting of the
normal bundle by $SU(2)_R$ is compatible with $G_2$ holonomy.}

For singularities of type $D$ or $E$, the twisting by $\Lambda$ has been
studied in the context of Calabi-Yau compactification and
 plays an important role \nref\aspgro{P.
Aspinwall and M.
Gross, ``The $SO(32)$ Heterotic String On A K3 Surface,'' hep-th/9605131.}%
\nref\vkone{M. Bershadsky, K. Intriligator, S. Kachru, D. R.
Morrison, V. Sadov, and C. Vafa, ``Geometric Singularities And
Enhanced Gauge Symmetries,'' Nucl. Phys. {\bf B481} (1996) 215.}%
\refs{\aspgro,\vkone}. In these examples, $\Gamma$ is a nonabelian
group and $\Lambda$ is a finite group which can be identified with
the group of outer automorphisms of the $D$ or $E$ gauge group.
Twisting by $\Lambda$ means, in this case, that the $D$ or $E$
gauge theory can be twisted, as one goes around a
non-contractible one-cycle, by an outer automorphism.
On either a Calabi-Yau manifold or a manifold of $G_2$ holonomy,
this can give a way to break the gauge group to a non-simply-laced subgroup.

We want to focus here on the case of a singularity of type $A$,
so that the gauge group is $SU(N)$ for some $N$.  In this case,
$\Gamma$ is a $\Z_N$ subgroup of $SU(2)_R$ that we can take to
consist of matrices of the form \eqn\bino{\left(\matrix{ e^{2\pi
i k/N} & 0 \cr 0 & e^{-2\pi ik/N}}\right).} If such a matrix
acts on a column vector $\left(\matrix{a\cr b\cr}\right)$, then
the $\Gamma$-invariants are $x=a^N$, $y=b^{-N}$, and $z=ab$,
obeying the familiar equation \eqn\dino{xy=z^N} of the $SU(N)$
singularity.

The group of $SU(2)_R$ matrices that map $\Gamma$ to itself is in this case
 $\Lambda=O(2)$. Here $O(2)$ is generated by a discrete $\Z_2$
symmetry that exchanges $a$ and $b$, and a $U(1)$ subgroup of
diagonal matrices.  The $\Z_2$ symmetry corresponds to an outer
automorphism (complex conjugation) of $SU(N)$, analogous to the
discrete symmetries for the $D$ and $E$ groups; the associated
physics is similar. We want to focus on the continuous group
$\Lambda'\cong U(1)$.  It consists of matrices
\eqn\vino{\left(\matrix{e^{ i\psi/N}& 0 \cr
                                0     & e^{- i\psi/N}\cr}\right),~~0\leq\psi\leq
                                          2\pi.}
Thus, the action on the invariants $x,y,z$ is
\eqn\gino{(x,y,z)\to (e^{i\psi}x,e^{-i\psi }y,z).}

Now consider $M$-theory on an eleven-manifold $Z$ with a family
of $SU(N)$ singularities on a codimension four submanifold $B$
(in our application, $Z=\R^4\times X$ and $B=\R^4\times Q$). The
normal space to $B$ might be twisted by $\Lambda'$.  (There could
be a more general twisting by disconnected elements of $\Lambda$,
but we do not wish to consider that case.)  In view of \gino, this
means that the normal space to $B$ can be described by
coordinates $x,y, $ $z$ that obey $xy=z^N$; however, they are not
functions but sections of certain line bundles over $B$.  In fact,
they are sections respectively of ${\cal L}$, ${\cal L}^{-1}$,
and ${\cal O}$, where ${\cal O}$ is a trivial line bundle and
${\cal L}$ is an arbitrary line bundle that incorporates the
twisting by $\Lambda'$.

In $M$-theory, ${\cal L}$ is not merely an abstract complex line
bundle; the metric on $Z$ induces a connection on ${\cal L}$.
Let $K$ be the curvature of this connection; then $K/2\pi$
represents the first Chern class of ${\cal L}$ and so has integer
periods. There is also an $SU(N)$ gauge field $A$ on $B$, with
curvature $F$; let $\omega_5(A) $ be the Chern-Simons five-form
of $A$ (normalized so its periods are gauge-invariant mod
$2\pi$).  We claim that in the long wavelength limit of
$M$-theory on $Z$, there is an interaction of the form
\eqn\unic{I=\int_B {K\over 2\pi}\wedge \omega_5(A).} We will
first explore the consequences of this assumption and then show
that the interaction must be present.

$SU(N)$ gauge-invariance of $I$ is proved by observing that under
an infinitesimal gauge transformation $A\to A-d_A\epsilon$,
$\omega_5(A)$ transforms by addition of an exact form, a multiple
of $d\,\tr \,\epsilon F\wedge F$; upon integrating by parts and
using the fact that $dK=0$, this suffices to prove invariance of
$I$ under infinitesimal gauge transformations. We must also
consider the behavior under disconnected gauge transformations;
since $\omega_5(A)$ has periods that are gauge-invariant  mod
$2\pi$, \unic\ has been normalized so that $I$ is invariant mod
$2\pi$  under disconnected gauge transformations, which is good
enough for quantum field theory. Thus, a coefficient multiplying
the right hand side of \unic\ must be an integer; the discussion
below will show that (with the right choice of orientations) this
integer is 1.

Next, let us specialize to $Z=\R^4\times X$, $B=\R^4\times Q$ and
explore the consequences of having an interaction of the form of
\unic. We suppose that there are finitely many points
$P_\alpha\in Q$ at which the singularity of the normal space to
$Q$ is worse than an $SU(N)$ singularity, so that the line bundle
${\cal L}$ is not defined.  Away from these points, $K$ is
defined and obeys $dK=0$, but there might be delta function
contributions at the $P_\alpha$: \eqn\jucov{dK=2\pi \sum_\alpha
n_\alpha \delta_{P_\alpha}.} Here $\delta_{P_\alpha}$ is a delta
function supported at $P_\alpha$.  The $n_\alpha$ are integers
because the period of $K$, integrated over a small surface
$S_\alpha\subset Q$ that wraps around $P_\alpha$, is an integer
multiple of $2\pi$; in fact, the restriction of ${\cal L}$ to
$S_\alpha$ has first Chern class $n_\alpha$.  By integrating
\jucov\ over $Q$, we learn that \eqn\huxo{\sum_\alpha n_\alpha =
0.}

Is the interaction $I$ gauge-invariant when $n_\alpha\not= 0$?
Just as in section 2, in the presence of the singularity, we get
an anomaly under gauge transformations.  In fact, under $A\to
A-d_A\epsilon$, with $\omega_5$ shifted by a multiple of
$d\tr\,\epsilon F\wedge F$, integration by parts shows that the
change of $I$ under an infinitesimal gauge transformation is
\eqn\komiko{\delta I = -\sum_\alpha n_\alpha\int_{\R^4\times
P_\alpha} \tr\,\epsilon \,\,{F\wedge F\over 8 \pi^2}.} Hence,
gauge-invariance is only maintained if at each $P_\alpha$, there
are charged chiral multiplets (or more exotic degrees of freedom)
with an $SU(N)^3$ anomaly $n_\alpha$. Rather as in section 2,
\huxo\ is the condition for anomaly cancellation in the effective
four-dimensional theory.

For $D$ and $E$ singularities, $\Lambda$ is a finite group and we
do not get such a mechanism for anomalies in the four-dimensional
theory.  Such a mechanism is not needed, since in any event  the
$D$ and $E$ groups admit no anomalies in four dimensions, as their
Lie algebras have no cubic symmetric invariant.

\bigskip\noindent{\it An Example}

To show now that the interaction \unic\ is really present, it
suffices to show independently in one special case that charged
degrees of freedom with anomaly $n_\alpha$ are really present.
For this, we consider the example of a cone on a weighted
projective space $Y={\bf WCP}^3_{N,N,1,1}$ as considered in
section 3.7 of \aw.  We describe the weighted projective space by
homogeneous complex coordinates $(u_1,u_2,u_3,u_4)$, not all
zero, modulo \eqn\hdun{(u_1,u_2,u_3,u_4)\to
(\lambda^Nu_1,\lambda^Nu_2,\lambda u_3,\lambda u_4)
,\,\,\lambda\not=0.} $Y$ has $\Z_N$ orbifold singularities on the
locus $u_3=u_4=0$, which is a copy $U$ of $\CP^1=\S^2$.  A cone
$X$ on the weighted projective space $Y$ is constructed by
imposing the equivalence relation \hdun\ on the $u_i$ only for
$|\lambda|=1$.

The locus $Q$ of singularities in $X$ is a cone on $U=\S^2$.  A
cone on $\S^2$ is $\R^3$. So $Q=\R^3$ and in particular is
smooth. The generic singularity of $X$ is a $\Z_N$ orbifold
singularity, but at the ``origin,'' the  apex $P$ of the cone,
the singularity is worse.  Though smooth, $Q$  passes through
$P$.  The fact that the codimension-four submanifold $Q$ passes
through the isolated (non-orbifold) singularity $P$ has a
topological explanation: it occurs because the two-sphere $U$
wraps a non-trivial cycle in $Y$, and hence $Q$ cannot be slipped
away from the singularity.

Now let us check that (away from $P$) the normal bundle
to $Q$ is an $SU(N)$ singularity.  The normal coordinates to $U$ (or $Q$)
are $u_3$ and $u_4$, but subject to the orbifolding group
$(u_3,u_4)\to (\lambda u_3,\lambda u_4)$, where now (to get trivial action
on $u_1$ and $u_2$) $\lambda$ is an $N^{th}$ root of unity.
Clearly, if we set $a= u_3$,  $b=\overline u_4$,
this coincides with the description of the $SU(N)$ singularity in \bino.

Now we can determine the line bundle ${\cal L}$. We rewrite
\hdun\ in the form \eqn\jorry{(u_1,u_2,u_3,u_4)\to (t u_1,t u_2,
t^{1/N} u_3, t^{1/N} u_4)} with $t=\lambda^{1/N}$.  $U={\bf
CP}^1$ is defined by $u_3=u_4=0$ and has $u_3$ and $u_4$ as
normal coordinates.  Actually, $u_3$ and $u_4$ are sections of a
``line bundle'' over $U$.  Because of the exponent $1/N$ in
\jorry, this ``line bundle'' has first Chern class $1/N$ (and so
the ``line bundle'' must be defined in an orbifold sense).
However, the invariants $x=a^N=u_3^N$ and $y=b^N=\overline u_4^N$
transform with exponents $\pm 1$, and so are functions on ${\cal
L}^{\pm 1}$, where ${\cal L}$ is an ordinary line bundle over $U$
with first Chern class 1. So the integer $n$ associated with this
particular singularity is 1.

Hence we expect that charged degrees of freedom with an $SU(N)^3$ anomaly
of 1 will be present at this singularity.  This agrees with the analysis in section
3.7 of \aw, where it was argued that this type of singularity supports
a chiral multiplet in the fundamental representation of $SU(N)$.

The interested reader can analyze in a similar fashion the more general
weighted projective space ${\bf WCP}^3_{N,N,M,M}$ which was considered in
\aw.

\bigskip\noindent{\it Inclusion Of Abelian Gauge Symmetries}

Now let us look at $X$ a little more globally.  Away from a
finite set of points $P_\alpha$, it is an orbifold. Near
$P_\alpha$, $X$ looks like a cone on some six-dimensional
orbifold $Y_\alpha$.  We can proceed just as in section 2 to
analyze unbroken abelian gauge symmetries that come from the
$M$-theory three-form field $C$.  As in section 2, we let $X'$ be
a manifold-with-boundary obtained by omitting  small
neighborhoods of the $P_\alpha$.  The boundary of $X'$ is the
union of the $P_\alpha$.  We propose that the abelian gauge group
from the $C$-field has  Lie algebra $H^2(X';\R)$ just as in
section 2.\foot{$H^2(X';\R)$ is defined as the ordinary de Rham
cohomology of the orbifold.  We do not include any ``twisted
sector'' degrees of freedom at the fixed points; they have
already been included in the nonabelian gauge symmetry associated
with the orbifold.  We will not try to analyze the integral
structure of $H^2(X')$ or to determine the global form of the
gauge group.} As in section 2, we let $w_1,\dots, w_r$,
$r=b_2(X')$, be a basis of harmonic forms on $X'$.

The $U(1)^3$ anomalies can be treated just as in section 2; since
we worked only at the level of differential forms in section 2,
the analysis is unchanged by the fact that $X'$ is an orbifold.
We now want to analyze the $U(1)\cdot SU(N)^2$ anomalies. (The
following discussion actually applies for all the $A-D-E$ groups,
not just $SU(N)$.)   For this, the key point is the existence of a
(standard) interaction \eqn\ungin{I=\int_B C\wedge {\tr\,F\wedge
F\over 8\pi ^2},} in the long wavelength limit of $M$-theory,
away from singularities. Here $\tr\, F\wedge F/8\pi^2$ is the
(normalized) $SU(N)$ instanton number. This interaction is
invariant under a gauge transformation $C\to C+d\epsilon$ if the
orbifold locus $B$ is smooth. When $B$ has singularities, we meet
anomalies just as in section 2 and above.

In our case, $B=\R^4\times Q$. In using the low energy
interaction \ungin, we must excise small neighborhoods of the
singular points $P_\alpha$
and integrate only over $\R^4\times Q'$ where $Q'$ is $Q$ with these
neighborhoods removed; we denote the boundaries of these
neighborhoods as $U_\alpha$. Transforming $C\to C+d\epsilon$ and
integrating by parts, we find that the change in $I$ is a sum of
local contributions at the $P_\alpha$: \eqn\bungin{\delta_\alpha
I=-\int_{\R^4\times U_\alpha}\epsilon \wedge {\tr F\wedge F\over
8\pi^2}.} Just as in section 2, to express this in the low energy
theory, we consider the case that $\epsilon=\sum_{i=1}^r
\epsilon^{(i)} w_i$, with $\epsilon^{(i)}$ being functions on
$\R^4$. The local anomaly is \eqn\tungin{\delta_\alpha I=-\sum_i
\int_{\R^4}\epsilon^{(i)}{\tr\, F\wedge F \over 8\pi^2} \cdot
\int_{U_\alpha} w_i.} To cancel this anomaly, the chiral degrees
of freedom at $\R^4\times P_\alpha$ must have a $U(1)_i\cdot
SU(N)^2$ anomaly (here $U(1)_i$ is the $i^{th}$ copy of $U(1)$ in
the gauge group, generated by the harmonic form $w_i$) equal to
$\int_{U_{\alpha}} w_i$.  Now we can demonstrate anomaly
cancellation in the effective four-dimensional theory; rather as
in the other examples we have considered, we merely note that
\eqn\hubok{\sum_\alpha\int_{U_\alpha} w_i=\int_{Q'}dw_i=0.} So
the $U(1)_i\cdot SU(N)^2$ anomalies of the chiral fields on the
various singularities add up to zero.

We can again illustrate this with the example of the cone on $Y=
{\bf WCP}^3_{N,N,1,1}$.  The second Betti number of $Y$ is $1$,
so there is a single $U(1)$ to consider, generated by a harmonic
two-form $w$ on $Y$.  Such a $w$ has   $\int_U w \not= 0$; here
$U$, the locus of orbifold singularities in $Y$, is a copy of
${\bf CP}^1$ as seen above. So the chiral $SU(N)$ degrees of
freedom found above must be charged under the $U(1)$. This agrees
with the result in \aw, where (using the fact that the cone on
$Y$ is dual to a configuration of intersecting branes in $\R^6$)
it was seen that the global form of the gauge group is $U(N)$, not
$SU(N)\times U(1)$, and that the chiral degrees of freedom are in
the fundamental representation of $U(N)$.

\bigskip
This work was supported in part by NSF Grant PHY-0070928.  I
would like to thank K. Intriligator and B. Acharya for
discussions. \listrefs
\end